\documentclass[sigplan,10pt]{acmart}

\renewcommand\footnotetextcopyrightpermission[1]{}
\usepackage{xspace}
\usepackage{algorithm}
\usepackage{algpseudocode}

\newcommand{\system}{Xema\xspace}

\graphicspath{{figs/}}

\begin{document}

\title{Xema: Efficient Diffusion Serving through Fine-Grained Memory Management and Auto-Configuration}

\author{Xueze Kang}
\authornote{Equal contribution.}
\affiliation{%
  \institution{The Hong Kong University of Science and Technology (Guangzhou)}
  \city{Guangzhou}
  \country{China}}
\email{xkang507@connect.hkust-gz.edu.cn}

\author{Guangyu Xiang}
\authornotemark[1]
\affiliation{%
  \institution{The Hong Kong University of Science and Technology (Guangzhou)}
  \city{Guangzhou}
  \country{China}}
\email{gxiang190@connect.hkust-gz.edu.cn}

\author{Suyi Li}
\affiliation{%
  \institution{Hong Kong University of Science and Technology}
  \city{Hong Kong}
  \country{China}}
\email{slida@cse.ust.hk}

\author{Yuxin Wang}
\affiliation{%
  \institution{No institutional affiliation}
  \city{}
  \country{}}
\email{yxwang.ele@gmail.com}

\author{Shaohuai Shi}
\affiliation{%
  \institution{Harbin Institute of Technology, Shenzhen}
  \city{Shenzhen}
  \country{China}}
\email{shaohuais@hit.edu.cn}

\author{Lin Zhang}
\authornote{Corresponding author.}
\affiliation{%
  \institution{Hong Kong University of Science and Technology}
  \city{Hong Kong}
  \country{China}}
\email{lzhangbv@connect.ust.hk}

\author{Xiaowen Chu}
\authornotemark[2]
\affiliation{%
  \institution{The Hong Kong University of Science and Technology (Guangzhou)}
  \city{Guangzhou}
  \country{China}}
\email{xwchu@hkust-gz.edu.cn}

\renewcommand{\shortauthors}{Kang et al.}

\begin{abstract}
Diffusion models are increasingly deployed as production visual-generation services, where serving high-resolution image and long video generation is often limited by GPU memory. 
% Existing memory controls, such as weight offloading, sharding, and VAE slicing, are difficult to use efficiently because the dominant memory bottleneck varies across request templates, while coarse-grained controls often incur excessive performance overhead for memory reduction. 
Popular memory-saving techniques such as weight offloading, sharding, and VAE slicing are often not practical because they tend to introduce significant performance overhead.
In this paper, we present \system, a memory-efficient diffusion serving system that exploits predictable tensor lifetimes for trace-guided memory optimization. For each request template, \system derives an offline memory trace to identify short memory-pressure intervals and applies memory mitigation only within these intervals and only by the amount needed to fit the target GPU budget. \system further constructs a static memory layout for tensors with predictable lifetimes, reducing fragmentation-induced reserved memory and making offline memory reasoning reliable at runtime. Built on this memory optimization layer, \system introduces an offline planner that jointly selects parallelism, concurrency, and memory control under GPU memory and SLO constraints. The selected plan is stored in a plan table and directly used by the online serving runtime. We implement \system on production diffusion pipelines and evaluate it with Flux.2, CogVideoX-5B, and LTX-2. Compared with existing serving configurations, \system improves SLO attainment by up to 3.7$\times$, and reduces planning cost from 6.3~hours to 197~seconds compared with grid search.
\end{abstract}

\maketitle

\section{Introduction}
\label{sec:intro}

\begin{figure}[t]
\centering
\includegraphics[width=0.9\linewidth]{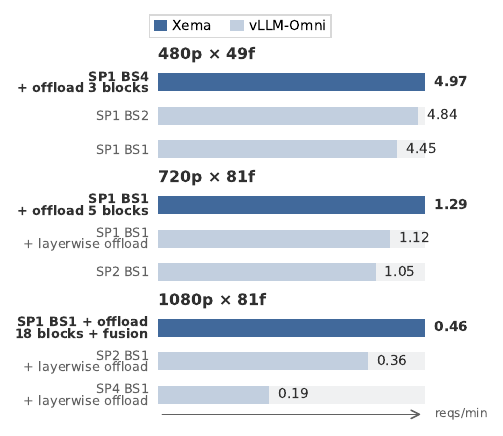}
\caption{Optimal execution configurations vary across LTX-2 request templates. For each template, the first bar is selected by
\system, and the next two bars are the best and second-best configurations searched by vLLM-Omni. By expanding the search space with fine-grained memory controls, \system finds higher-goodput configurations than vLLM-Omni.}
\label{fig:motivation-bestconfig}
\end{figure}

Diffusion models are now widely used in visual generation, including image generation~\cite{ldm,dit,glide,dalle2,imagen,pixartalpha}, image editing~\cite{sdedit,instructpix2pix,controlnet,dragdiffusion}, video generation~\cite{makeavideo,imagenvideo,cogvideox,ltxvideo}, 3D content creation~\cite{dreamfusion,magic3d}, and interactive design~\cite{play}. These applications are moving from offline demonstrations to production services, where the serving system must handle many user requests under latency and resource constraints. At the same time, production workloads are shifting toward more demanding request templates, such as high-resolution images and long videos. In this paper, a request template refers to a supported combination of model, output resolution, and frame count. These trends make efficient diffusion serving an important systems problem for deploying generative AI applications at scale.

However, GPU memory has become a major bottleneck in serving modern diffusion requests. To illustrate this problem, we measure the peak GPU memory of representative image and video requests. For example, generating a \(1024 \times 1024\) image with Flux.2 requires more than 64~GB of GPU memory, and generating a \(1080p \times 49f\) video with CogVideoX requires more than 100~GB, both with a batch size of one. The memory pressure also comes from different parts of the pipeline. Large image models can be dominated by model weights, while long-video requests can be dominated by activations. This request-dependent behavior makes memory control difficult as reducing the wrong part of the footprint may not help the request fit on a target GPU. Because serving can run only when reserved memory fits within the target GPU budget, these large and heterogeneous footprints directly determine whether a request template is serviceable. Under fixed production GPU resources, memory management is therefore necessary to make diverse image and video requests fit while preserving efficiency.

Current memory control mechanisms are insufficient for this setting. Weight offloading~\cite{vllmomni-cpu-offload} and sharding~\cite{vllmomni-hsdp} reduce the weight footprint, but they introduce CPU--GPU transfers and cross-GPU communication, respectively, and do little for activation-dominated requests. VAE (Variational Autoencoder) slicing~\cite{vllmomni-serve} can reduce VAE activation peaks, but it may slow down the VAE stage and does not address peaks from denoising. Moreover, their control granularity can be too coarse. For example, in Flux.2, offloading a subset of blocks is enough to fit a 48~GB GPU, but existing mechanisms only support full offloading, increasing per-step latency from 3.1~s to 8.4~s. Therefore, current systems lack request-aware selection of memory controls, and their coarse granularity often trades more performance than necessary for memory reduction.

The key opportunity is that diffusion inference is template-static and its memory pressure is localized. For a fixed request template, the execution path of diffusion inference is determined by the model, output resolution, frame count, and denoising steps. As a result, the operator sequence, tensor shapes, and tensor lifetimes can be derived before serving real inputs. Moreover, diffusion inference does not sustain peak memory throughout the whole request. Our measurements show that memory usage is highly uneven. In CogVideoX, only about 2\% of allocation events reach at least 90\% of the phase peak. This suggests that memory control does not need to be applied globally. Instead, it can focus on the few intervals where the offline-derived memory trace exceeds the GPU budget.

We present \system, a diffusion serving system that exploits template-static tensor lifetimes to support low-overhead memory mitigation. For each request template, \system derives an offline memory trace and identifies short memory-pressure intervals that may exceed the target GPU budget. At runtime, \system mitigates memory only within these intervals and only by the amount required to fit the GPU, avoiding unnecessary transfer, communication, or serialization overhead in regions that already fit. \system also builds a static memory layout for tensors, reducing fragmentation-induced reserved-memory overhead and ensuring that runtime allocation follows the offline-derived trace .

Fine-grained memory optimization makes large requests feasible, but it also creates a larger execution space that should be planned automatically. The serving system must decide not only which memory mitigation to apply, but also how it should interact with parallelism and concurrency configurations. These choices are tightly coupled. For instance, sequence parallelism can reduce per-GPU activation memory but introduces additional communication, while increasing batch size can improve utilization but raises activation memory. As a result, the best execution plan changes across request templates. Figure~\ref{fig:motivation-bestconfig} shows this effect for LTX-2: small, medium, and large video requests prefer different combinations of spatial parallelism, batch size, and memory mitigation configurations. Manually tuning these choices for every model, resolution, and frame count is impractical in production services. 

To address this challenge, \system builds an offline planner that jointly optimizes parallelism, concurrency, and memory control under GPU memory and SLO constraints. For each request template, the planner evaluates candidate execution configurations using the offline-derived memory trace. It prunes configurations that cannot satisfy the memory budget or SLO, focuses mitigation search on the intervals where memory may exceed the budget, and uses profiling results to select the highest-goodput feasible plan, where goodput is the throughput of requests that meet their SLOs~\cite{distserve}. The selected plan is stored in a plan table, so the online serving runtime only needs to look up the precomputed plan for each incoming request template. 

This paper makes the following contributions:

\begin{itemize}
\item We identify GPU memory as a major bottleneck in diffusion serving. We show that image and video requests can exceed GPU memory even at batch size one, and that the dominant memory source varies across request templates, including model weights, DiT and VAE activations. 

\item We design a trace-guided memory optimization layer for diffusion serving. It exploits template-static tensor lifetimes to locate short memory-pressure intervals, applies memory control only where and by the amount needed, and constructs a static memory layout to reduce memory fragmentation.

\item We design an offline auto-configuration planner that jointly selects parallelism, concurrency, and memory control under GPU memory and SLO constraints. The planner uses offline-derived memory traces to prune infeasible candidates, restrict mitigation search to memory-pressure intervals, and select the highest-goodput feasible plan for each request template.

\item We implement Xema on production diffusion pipelines and evaluate it on Flux.2, CogVideoX-5B, and LTX-2. Xema improves SLO attainment by up to 3.7$\times$, expands the range of request templates that can be served under limited GPU memory, and reduces planning cost compared with exhaustive search.
\end{itemize}

\section{Background}
\label{sec:background}

\subsection{Diffusion Inference Pipeline}

Diffusion inference consists of three stages: a text encoder, a denoiser, and a VAE decoder. The text encoder converts the user prompt into conditioning embeddings and is executed once at the beginning of a request. 

The denoiser then iteratively refines a latent representation over multiple denoising steps. Modern image and video diffusion models commonly use a Diffusion Transformer (DiT) denoiser composed of repeated transformer blocks. This stage usually dominates inference computation.

After denoising, the VAE decoder maps the final latent representation back to pixels, producing the output image or video frames. The VAE decoder is executed once at the end of the request and is commonly built from convolution, normalization, residual, and upsampling modules. Although the VAE decoder runs only once, it can still create large activation peaks for high-resolution outputs because it operates close to the final spatial resolution.

Image and video generation services usually support a finite set of request specifications, such as model type, output resolution, and video length. We refer to each supported specification as a request template and denote it as
\[
r = (model, H, W, T),
\]
where \(model\) is the diffusion model, \(H\) and \(W\) are the output height and width, and \(T\) is the number of output frames. For image generation, \(T = 1\). Requests with different templates execute the same high-level pipeline, but their tensor shapes and per-stage workloads differ according to the requested resolution and frame count.

\subsection{Memory Overhead in Diffusion Inference}

The GPU memory footprint of diffusion inference mainly comes from four sources: model weights, DiT activations, VAE activations, and reserved-memory overhead from the runtime allocator. These sources have different scaling behaviors across models and request templates.

\textbf{Model weights.}
Model weights include the parameters of the text encoder, DiT denoiser, and VAE decoder. For a fixed model and precision, this footprint is independent of output resolution and frame count, but it can still dominate memory for large models because modern diffusion models usually contain tens of billions of parameters. For example, Flux.2 has 32B of parameters.

\textbf{DiT activations.}
The DiT denoiser materializes activations and operator intermediates from attention and feed-forward blocks at each denoising step. Their footprint grows with the latent sequence length, which scales approximately with \(H \times W \times T\), making DiT activations a dominant memory source for long-video generation.

\textbf{VAE activations.}
The VAE decoder maps the final latent representation back to pixels. For videos, VAE decoding is usually performed sequentially over frames or small temporal chunks rather than over the entire video at once, so its peak activation footprint is mainly determined by spatial resolution and the frames decoded per call, scaling approximately with \(H \times W\). Since decoding operates near the output resolution, high-resolution outputs can still create substantial memory pressure.

\textbf{Reserved-memory overhead.}
In addition to live tensors, GPU runtimes reserve memory through caching allocators. We distinguish \emph{active memory}, occupied by live tensors, from \emph{reserved memory}, held by the allocator. Reserved memory can exceed active memory because freed blocks may not match the size requirements of later allocations, creating allocator overhead and fragmentation. This distinction matters because an execution must fit its reserved memory, not only its active memory, within GPU capacity.
\subsection{Execution Configurations}

Diffusion serving systems expose several execution configurations that determine how inference is mapped to available GPUs and how requests are executed. We summarize three common dimensions: parallelism, concurrency, and memory control.

\textbf{Parallelism.}
Parallelism controls how inference is distributed across GPUs. Data parallelism (DP) assigns different requests to different GPUs or GPU groups, sequence parallelism (SP) partitions the latent sequence across GPUs, and hybrid sharded data parallelism (HSDP) shards model parameters and gathers them during execution These parallel modes affect request-level concurrency, activation memory, and weight memory in different ways.

\textbf{Concurrency.}
Concurrency specifies how many requests are executed together, typically represented by batch size in diffusion serving. Requests are usually batched only within the same request template because tensor shapes and execution schedules differ across templates; larger batches improve GPU utilization but increase resident activations and intermediates.

\textbf{Memory control.}
Memory control refers to execution techniques that reduce GPU memory usage. Existing diffusion serving systems commonly provide weight offloading and VAE slicing. Weight offloading stores part of the model weights in CPU memory and transfers them back to GPU when they are needed. VAE slicing reduces VAE memory usage by processing the VAE computation in smaller slices rather than materializing the full computation at once. There are also \emph{lossy} approaches for reducing memory usage, such as quantization and VAE tiling. These lossy techniques are complementary to our work, which focuses on lossless system-level optimizations. 

\section{Motivation}
\label{sec:motivation}

\subsection{GPU Memory Is the Bottleneck}
\label{subsec:motivation-1}

The memory pressure of modern diffusion inference is rapidly increasing. As image generation moves to higher resolutions and video generation moves to longer durations, even a single request can exceed the capacity of common GPUs, as discussed in the introduction. Therefore, GPU memory capacity already limits which request templates can be served before considering higher concurrency or throughput optimization.

We further dive into the memory breakdown and find that the memory bottleneck differs across request templates. First, the bottleneck may come from either model weights or activations. Image generation can be weight-dominated because the latent sequence is relatively short while the model parameters are large. For example, in Flux.2 \(1024^2\) image generation, weights occupy 60.2 GB while activations occupy 4.1 GB. In contrast, video generation can be activation-dominated because the latent sequence grows with both spatial resolution and frame count. For CogVideoX \(1080p \times 49f\), weights occupy only 10.8 GB, but activations occupy 89.6 GB.

Second, activation peaks may appear in different stages of the diffusion pipeline. During DiT denoising, intermediate tensors inside transformer blocks are repeatedly materialized across denoising steps and can form large per-block peaks. During VAE decoding, high-resolution feature maps produced by convolution and upsampling modules can also create substantial peaks, even though the VAE decoder is executed only once at the end of the request. Thus, as shown in memory trace from Figure~\ref{fig:memory-cdf}, the peak activation footprint may be determined by either the repeated DiT stage or the final VAE stage, depending on the request template.

Third, the actual memory requirement can be further amplified by allocator fragmentation. The active memory occupied by live tensors can be lower than the memory reserved by the runtime allocator, because freed blocks may not match later allocation requests. In our measurements, the PyTorch caching allocator achieves only 58\%--89\% memory utilization on diffusion workloads. As a result, an execution may run out of memory even when its active tensor footprint appears to fit the GPU capacity.

\subsection{Existing Memory Controls Are Insufficient}

Existing diffusion serving systems provide several memory-control mechanisms, but they do not fully address the heterogeneous memory bottlenecks discussed in Section~\ref{subsec:motivation-1}.

First, existing controls are often too coarse-grained. Weight offloading and HSDP reduce GPU weight memory by moving or sharding model parameters, but they introduce CPU--GPU transfers or cross-GPU communication. Since they are commonly applied as large-scope switches, they may reduce more memory than necessary and add avoidable latency. For example, in Flux.2, offloading \(16\) blocks is already sufficient to fit a 48 GB GPU, while full offloading with \(56\) blocks increases per-step latency from 3.1 s to 8.4 s without improving goodput. VAE slicing has a similar issue: it reduces VAE activation memory by slicing the VAE computation, but it usually slows down the whole VAE stage even if only a short interval exceeds the memory budget.

Second, existing controls do not cover all important memory sources. Weight offloading and HSDP mainly target model weights, while VAE slicing targets VAE activations. However, large video requests can be dominated by DiT activations and operator intermediates during denoising. In such cases, existing mechanisms may still fail even when all of them are enabled. For example, vLLM-Omni still OOMs when serving CogVideoX \(1080p \times 289f\) on four 48 GB A6000 GPUs with all existing memory controls enabled.

Third, allocator-level mechanisms only partially address reserved-memory overhead. PyTorch's \texttt{expandable\_segments} can reduce fragmentation by mapping physical pages into a larger virtual address range, but it remains a runtime allocation policy. Its reserved-memory peak can still exceed the active tensor peak, and page-level mapping can add overhead for large requests with frequent tensor allocations.

\subsection{Memory Pressure Is Predictable and Localized}

\begin{figure}[t]
  \centering
  \includegraphics[width=\linewidth]{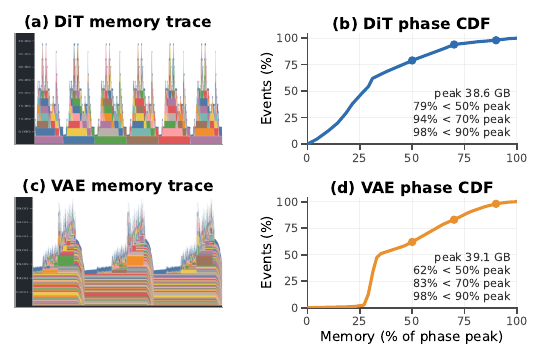}
  \caption{Memory usage is highly uneven during CogVideoX inference. Most allocation events are far below the phase peak, while only a small fraction of events reach near-peak memory usage.}
  \label{fig:memory-cdf}
\end{figure}

Although diffusion inference creates severe memory pressure, its memory behavior has two useful properties. First, diffusion inference has \emph{template-static execution}. For a fixed request template \(r=(model,H,W,T)\) and a fixed number of denoising steps, the operator sequence, tensor shapes, and allocation/free events are determined by the model architecture and request specification. They do not depend on the concrete content. Therefore, a lightweight shape-only execution can derive tensor lifetimes and construct the memory trace for a request template before serving real inputs.

Second, memory pressure is highly localized over the request lifecycle. Diffusion inference does not stay near its peak memory usage throughout execution. Instead, large memory demand appears only in short intervals. Figure~\ref{fig:memory-cdf} shows the memory-usage CDF of CogVideoX \(720p \times 289f\). In the DiT phase, 94\% of allocation events are below 70\% of the phase peak; in the VAE phase, 83\% of allocation events are below 70\% of the phase peak. In both phases, only about 2\% of events reach at least 90\% of the peak. 

These properties create an opportunity for more targeted memory optimization. Since the memory trace can be predicted for each request template, the system can identify the intervals where memory may exceed the GPU budget before execution. Since those intervals are short, memory reduction does not need to be applied globally across the whole model or the entire request lifecycle. We only need to reduce memory at the few intervals where the predicted memory trace exceeds the budget. 

\subsection{Configuration Planning Remains Necessary}

The predictability and locality of memory pressure create an opportunity for fine-grained memory control, but they do not remove the need for configuration planning. Once the system has multiple memory-control primitives, it must decide how to combine them with parallelism and concurrency. These choices are tightly coupled: increasing batch size can improve GPU utilization but also increases activation memory; increasing SP can reduce per-GPU activation memory but introduces communication; HSDP can reduce weight memory but requires weight-gather communication; memory mitigation can avoid OOM but may increase latency.

The best combination also varies across request templates. Figure~\ref{fig:motivation-bestconfig} shows that LTX-2 requires different highest-goodput configurations under different resolutions and frame counts. A small \(480p \times 49f\) request benefits from lower SP and higher batch size, while larger \(1080p\) requests require higher SP and, for longer videos, additional memory mitigation. Therefore, a fixed rule or manually tuned configuration cannot consistently achieve high goodput across templates. It motivates an offline planner that selects, for each request template, the execution configuration that satisfies the memory budget and SLO while maximizing goodput.
\section{System Overview}
\label{sec:overview}

\begin{figure}[t]
\centering
\includegraphics[width=\linewidth]{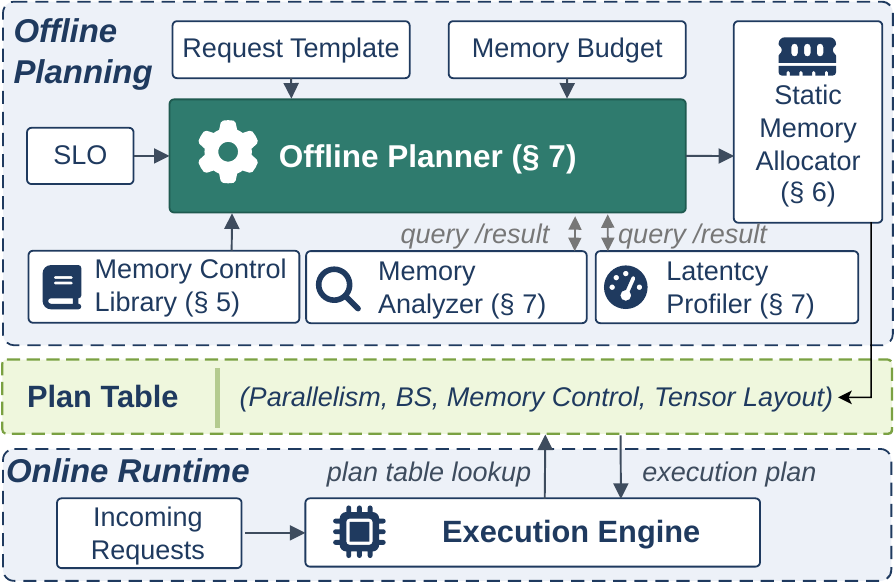}
\caption{System architecture.}
\label{fig:arch}
\end{figure}

\system is a diffusion serving system that exploits template-static tensor lifetimes to support low-overhead memory mitigation and automatic execution-configuration planning under GPU memory and SLO constraints.

Using the offline-derived memory trace, \system identifies memory-pressure intervals where execution would exceed the GPU budget. It then applies mitigation only to these intervals and only at the intensity needed to fit the budget. This fine-grained control lowers the performance tax of memory mitigation compared with globally enabling conservative memory controls. To cover different sources of memory pressure, \system also extends the memory-control library with primitives such as activation offloading, chunking, and fusion (\S\ref{sec:mechanisms}).

\system further uses a static memory allocator. By assigning address ranges according to tensor lifetimes, the allocator allows non-overlapping tensors to reuse the same memory region and reduces fragmentation-induced reserved memory. Because managed tensors are allocated at their precomputed offsets, the offline memory trace and the runtime memory trace are identical for these tensors (\S\ref{sec:layout}).

To maximize serving goodput under memory and SLO constraints, \system uses an offline planner to select execution configurations for each request template. The planner jointly considers memory control, parallelism, and concurrency; uses memory traces to prune infeasible candidates and restrict memory-control search to memory-pressure intervals; and uses profiling results to select the highest-goodput feasible plan (\S\ref{sec:planner}).

Figure~\ref{fig:arch} shows the architecture of \system. In the offline stage, the Memory Analyzer derives template-specific tensor lifetimes and memory traces. The Memory Control Library provides primitives for reducing memory pressure. The Static Memory Allocator produces precomputed tensor placements, and the Profiler measures the latency cost of candidate execution configurations. The Planner combines these outputs to generate a plan table for each request template and hardware configuration. In the online stage, the Serving Runtime maps each request to its template, looks up the selected plan, and executes the planned parallelism, batch size, memory-control actions, and static allocation.

\section{Fine-Grained Memory Mitigation}
\label{sec:mechanisms}

Existing systems often apply memory mitigation coarsely. Once a mechanism is enabled, it is applied globally across the request lifecycle. For example, vLLM-Omni's layerwise offloading keeps all non-executing layers on CPU throughout inference. For an execution configuration, memory only needs to stay below the budget; extra headroom does not improve goodput and often adds overhead. \system therefore refines memory mitigation along two dimensions: applying only the minimum amount needed, and applying it only where memory may exceed the budget.

\subsection{Applying Only as Much Mitigation as Needed}

\begin{figure}[t]
\centering
\includegraphics[width=\linewidth]{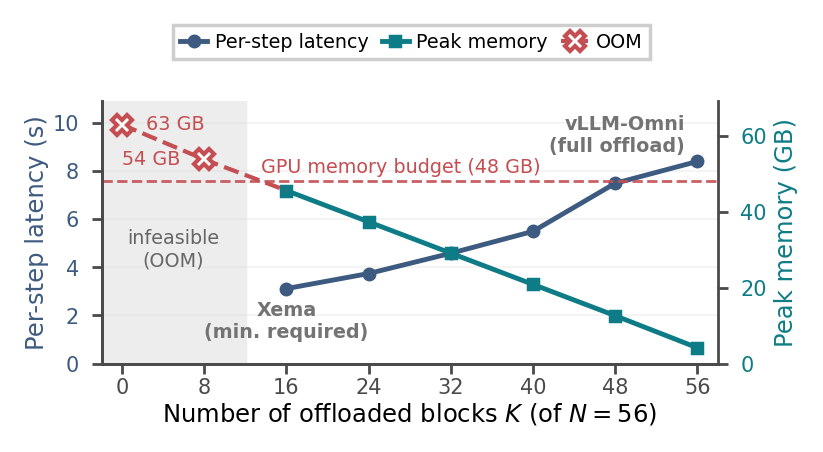}
\caption{Full offloading adds latency after memory fits the budget.}
\label{fig:offload-intensity}
\end{figure}

Existing systems often execute memory mitigation in full once a mechanism is enabled. We model a request as an execution trace $\mathcal{E}=(e_1,\ldots,e_n)$, where each trace segment $e_i$ is a maximal contiguous interval whose operator pattern remains unchanged. For a mitigation plan $\pi$, let $\mathbf{u}_i(\pi)$ be the mitigation action applied to segment $e_i$, and let $L_i(\cdot)$ and $M_i(\cdot)$ be the profiled latency and peak memory of that segment. Existing systems can only apply mitigation at full intensity, which may reduce memory more than needed and add unnecessary latency. \system instead applies only as much mitigation as needed by selecting
\[
\begin{aligned}
\pi^{*}
&=
\arg\min_{\pi}
\sum_{i=1}^{n}
\left(L_i(\mathbf{u}_i(\pi))-L_i(\mathbf{0})\right) \\
\text{s.t.}\quad
M_i(\mathbf{u}_i(\pi)) &\le B,\quad \forall i\in\{1,\ldots,n\}.
\end{aligned}
\]
Here, $\mathbf{0}$ denotes no mitigation, so the objective minimizes the latency overhead added by mitigation while satisfying the memory budget.

For transfer-based mitigation, this cost appears as communication on the critical path. Layerwise offloading must prefetch weights before each block, and HSDP must all-gather sharded weights before forward execution; when these transfers cannot be hidden by block computation, inference stalls. Figure~\ref{fig:offload-intensity} shows the tradeoff for block offloading in Flux.2. Increasing the number of offloaded blocks lowers peak memory but monotonically increases latency. The minimum feasible point offloads $K=16$ blocks and fits the 48GB budget, whereas full offloading ($K=56$) raises per-step latency from 3.1s to 8.4s without improving goodput.

\begin{figure}[t]
\centering
\includegraphics[width=\linewidth]{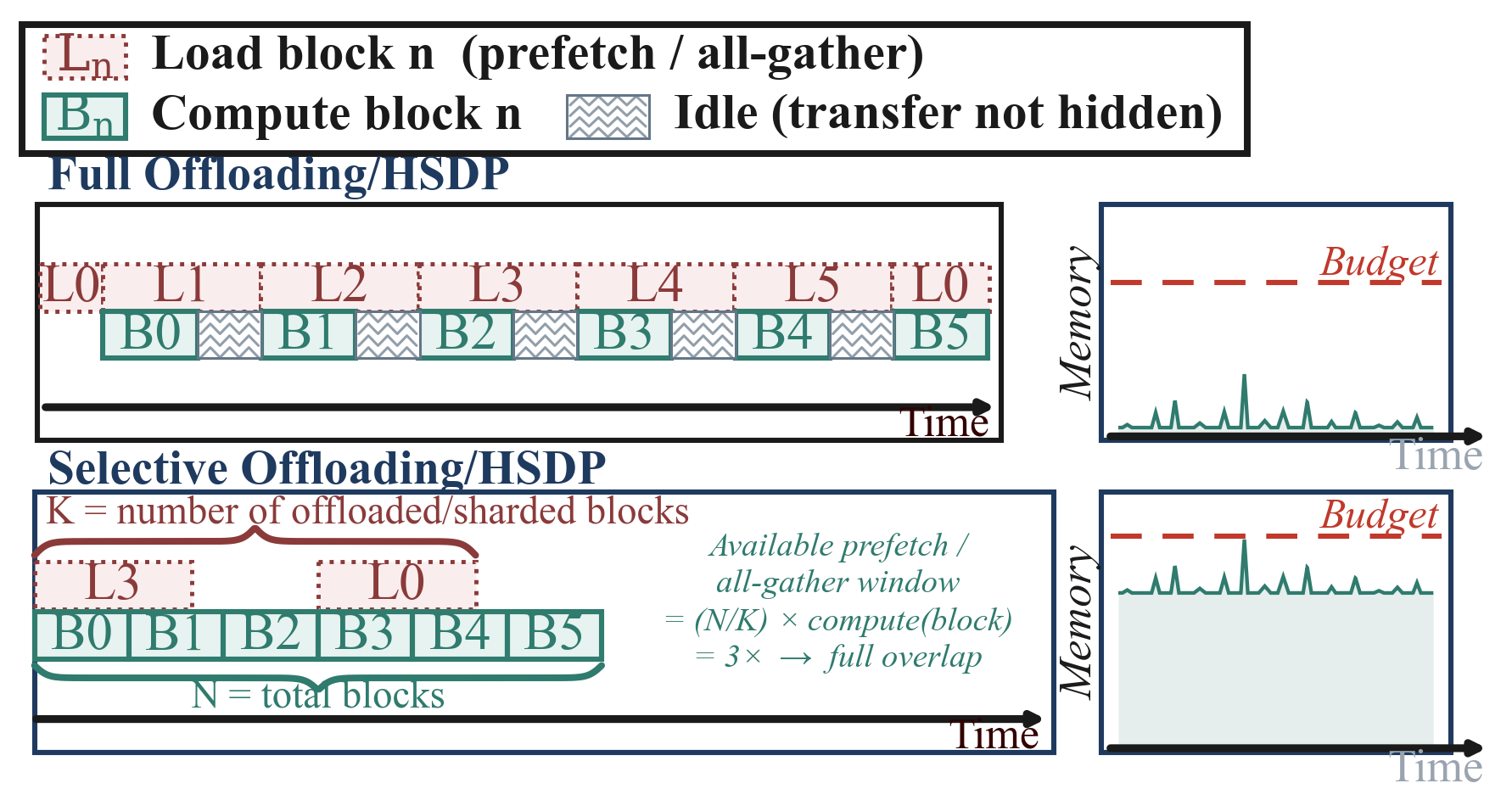}
\caption{Selective mitigation increases the overlap window for weight transfers.}
\label{fig:selective-mitigation}
\end{figure}

\textbf{Selective offloading and HSDP.}
\system applies only as much mitigation as needed at block granularity. \system exposes each weight block as an independent unit for weight-memory control. Given a memory budget, \system applies offloading or HSDP only to the subset of blocks needed to fit the budget, rather than applying to all blocks.

This selection reduces overhead in two ways. It reduces the amount of transferred data, and it increases the computation window available to hide each transfer. For a denoiser with $N$ DiT blocks, if $K$ mitigated blocks are selected uniformly, adjacent mitigated blocks are separated by about $N/K$ blocks. Thus, the overlap window grows from roughly one block under full mitigation to about $N/K$ blocks. Figure~\ref{fig:selective-mitigation} illustrates how this wider window makes CPU-GPU transfer or all-gather communication easier to overlap while still satisfying the memory budget.

\subsection{Applying Mitigation Only Where Needed}

Section~\ref{sec:motivation} shows that memory pressure appears only in short parts of the request lifecycle. Existing systems use one global mitigation setting for the entire request lifecycle, so mitigation also affects segments whose baseline memory already fits the budget. For mitigations that can be applied locally, this adds overhead without improving feasibility. We define the memory-pressure segments under the baseline execution as
\[
\mathcal{P}=\{i\mid M_i(\mathbf{0})>B\}.
\]
\system applies local mitigation only over $\mathcal{P}$ by selecting
\[
\begin{aligned}
\pi_{\mathcal{P}}^{*}
&=
\arg\min_{\pi}
\sum_{i\in\mathcal{P}}
\left(L_i(\mathbf{u}_i(\pi))-L_i(\mathbf{0})\right) \\
\text{s.t.}\quad
M_i(\mathbf{u}_i(\pi)) &\le B,\quad \forall i\in\mathcal{P},\\
\mathbf{u}_i(\pi) &= \mathbf{0},\quad \forall i\notin\mathcal{P}.
\end{aligned}
\]
Here, $\mathbf{0}$ denotes no mitigation, so the objective confines mitigation to memory-pressure segments and minimizes the resulting latency overhead while satisfying the memory budget. In our measurements, these memory-pressure segments cover only a small fraction of allocation events. Under CogVideoX-5B 1080p$\times$49f and LTX-2 1080p$\times$49f, only 3.70\% and 1.74\% of allocation events exceed the memory budget, respectively.

\textbf{Localized chunking and fusion.}
Chunking and fusion can reduce short-lived activation peaks, but they may add kernel overhead or restrict efficient operator implementations. \system therefore enables them only near operators whose execution may exceed the memory budget, such as DiT attention and VAE upsampling, and applies the minimum amount of chunking or fusion needed to fit the budget.

\begin{figure}[t]
\centering
\includegraphics[width=\linewidth]{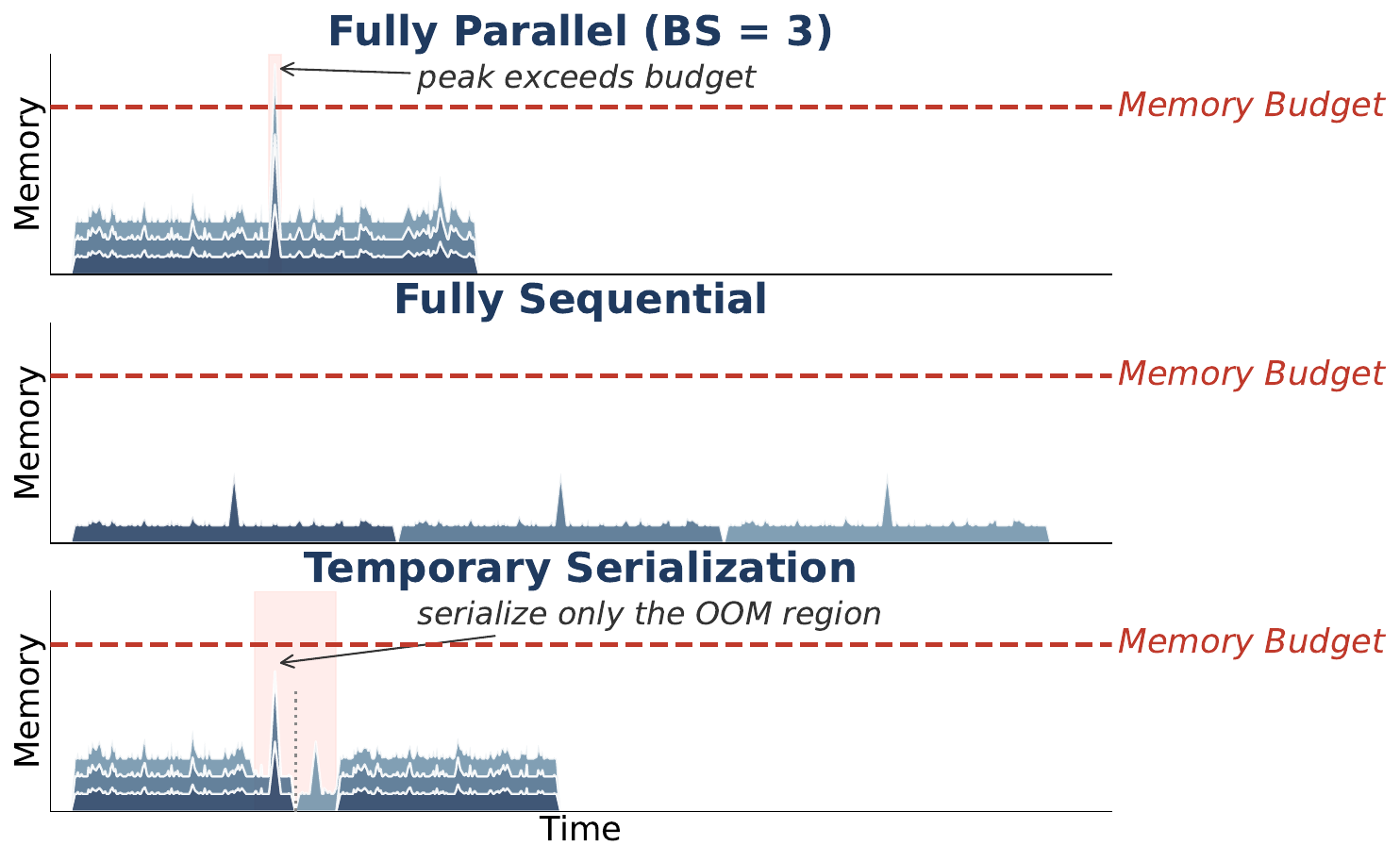}
\caption{Temporary serialization lowers concurrency only within memory-pressure intervals.}
\label{fig:temporary-serialization}
\end{figure}

\textbf{Temporary serialization.}
When batched requests exceed the memory budget only near a few peaks, reducing batch size globally wastes throughput. \system instead temporarily serializes part of the batch only at those peaks and restores the original batch size elsewhere. Figure~\ref{fig:temporary-serialization} illustrates this tradeoff: full sequential execution avoids OOM but serializes the whole lifecycle, whereas temporary serialization keeps most execution parallel while lowering concurrency only where memory may exceed the budget.

\section{Static Memory Layout}
\label{sec:layout}

PyTorch memory allocator assigns GPU memory addresses using only the current memory state. Because tensor allocation and release orders differ, freed blocks often do not match later tensor sizes, leaving holes that cannot be fully reused and increasing reserved memory beyond active memory. This fragmentation not only wastes available memory, but also leads to mismatch between offline-derived memory trace and runtime reserved-memory trace.

\textbf{Limitations of expandable segments.}PyTorch's \texttt{expand}\allowbreak\texttt{able\_segments} reduces fragmentation by reserving a contiguous virtual address range and mapping physical pages on demand, but it remains a runtime policy. The reserved peak still depends on execution-time allocation order, live blocks, and mapping decisions, so it cannot provide the offline memory bound required by \system. Expandable segments also incur runtime overhead from page-level memory mapping. PyTorch treats map/unmap operations as expensive, and larger requests tend to trigger these operations more often because they create larger and more frequent tensor allocations. This overhead can noticeably increase latency for high-resolution images or long videos. These limitations motivate an offline layout that uses template-static tensor lifetimes to assign addresses before execution.

\subsection{Offline Static Layout}

For each request template, most GPU memory is \emph{static}: model weights and graph-visible tensors have deterministic shapes and lifetimes across runs. The remaining \emph{dynamic} memory, such as CUDA contexts, NCCL buffers, and runtime-internal buffers, is not fully visible in the graph and depends on the runtime environment. Dynamic memory is typically only hundreds of MB in diffusion model serving, so \system can reserve a small margin for it and plan the dominant static portion offline.

\system obtains static tensor lifetimes by running the pipeline once with shape-only placeholder tensors. Data-independent control flow guarantees that this lightweight run produces the same allocation and release events as real inference.

\system then casts layout as rectangle packing on an address-time plane: each tensor occupies a rectangle with lifetime as width and size as height, and tensors with overlapping lifetimes must use disjoint address ranges. The objective is to minimize the highest used address, which determines the static reserved peak.

Algorithm~\ref{alg:static-layout} sorts tensors by size and uses best-fit placement for each tensor. Among the address gaps formed by tensors with overlapping lifetimes, it chooses the smallest gap that can hold the tensor; if no gap fits, it places the tensor above all conflicts. Placing large tensors first preserves large contiguous regions, while smaller tensors can fill the remaining gaps.

\begin{algorithm}[t]
\caption{Offline static memory layout}
\label{alg:static-layout}
\begin{algorithmic}[1]
\Require Tensor set $\mathcal{T}$
\Require Size $s_i$ and lifetime $[b_i,e_i)$ for each tensor $i$
\Ensure Address offset $a_i$ for each tensor $i$
\State Sort $\mathcal{T}$ in descending order of $s_i$
\State $\mathcal{P} \gets \emptyset$
\For{each tensor $i \in \mathcal{T}$}
  \State $\mathcal{O}_i \gets \{j \in \mathcal{P} \mid [b_i,e_i) \cap [b_j,e_j) \ne \emptyset\}$
  \State $\mathcal{G}_i \gets$ address gaps induced by $\mathcal{O}_i$
  \State $g \gets \arg\min_{g \in \mathcal{G}_i:\ |g|\ge s_i} |g|$
  \If{$g$ exists}
    \State $a_i \gets$ bottom address of $g$
  \Else
    \State $a_i \gets$ address above all tensors in $\mathcal{O}_i$
  \EndIf
  \State $\mathcal{P} \gets \mathcal{P} \cup \{i\}$
\EndFor
\State \Return $\{a_i\}$
\end{algorithmic}
\end{algorithm}

\begin{figure}[t]
\centering
\includegraphics[width=\linewidth]{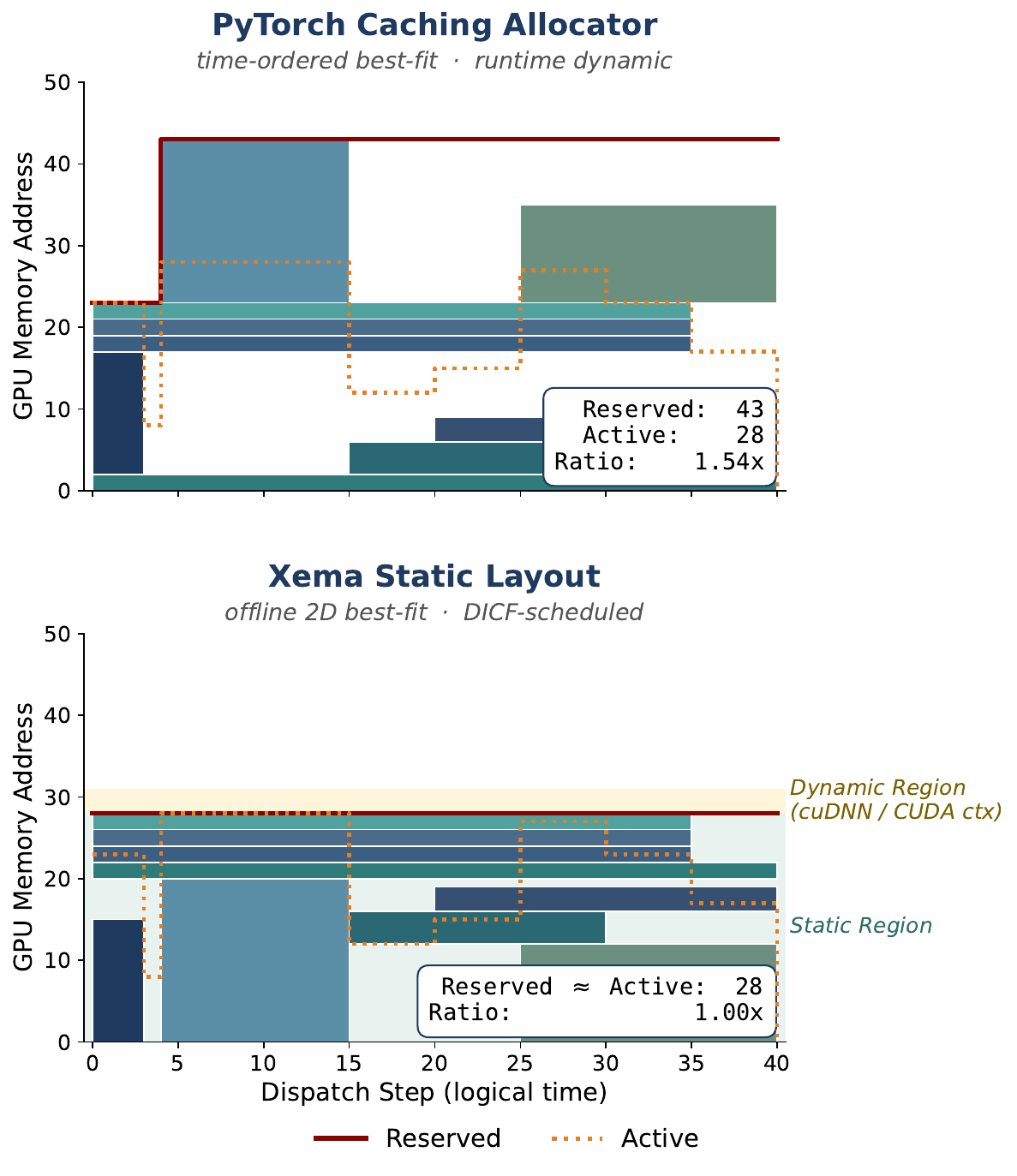}
\caption{\system reuses address ranges across non-overlapping tensor lifetimes.}
\label{fig:static-layout}
\end{figure}

Figure~\ref{fig:static-layout} contrasts this offline layout with PyTorch's online best-fit allocation. PyTorch only sees the current one-dimensional address space and cannot use future lifetime information, so reserved memory can exceed active memory. \system uses the full lifetime trace to reuse the same address range across tensors whose lifetimes do not overlap, making the reserved peak closely track active memory.

\subsection{Runtime Memory Management}

Static layout covers only graph-visible tensors. At runtime, \system divides GPU memory into two regions: the Static Region stores tensors at addresses determined by Algorithm~\ref{alg:static-layout}, while the Dynamic Region is a contiguous region above it and is managed by the PyTorch allocator. Since dynamic memory is small in diffusion model serving, \system reserves an additional 5\% memory margin beyond the static reserved peak for the Dynamic Region.

\section{Planner}
\label{sec:planner}

\begin{figure*}[t]
\centering
\includegraphics[width=\textwidth]{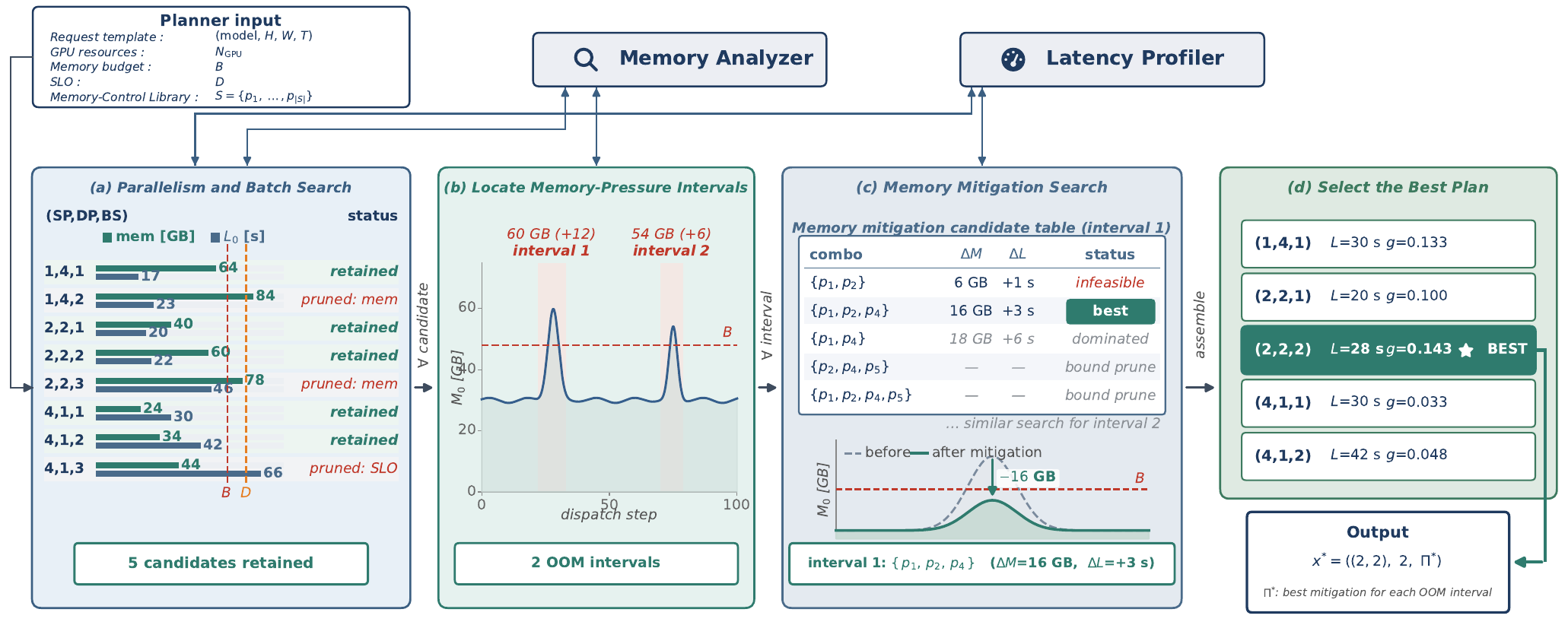}
\caption{\system planning workflow.}
\label{fig:planner-workflow}
\end{figure*}

After fine-grained memory mitigation expands the configuration space, \system must select an execution configuration that maximizes goodput under a memory budget and an SLO bound. Grid search is impractical because the search space spans parallelism layouts, batch sizes, memory control choices, and their fine-grained locations and intensities, and this search must be repeated for all request templates because their best configurations differ. As shown in Figure~\ref{fig:planner-workflow}, \system makes the search tractable by using memory traces to identify the lifecycle regions where memory demand may exceed the budget, and by restricting mitigation search in each region to the mechanisms that can reduce memory for the operators executed in that region.

\subsection{Problem Formulation}

For each request template $r=(model,H,W,T)$, the planner takes as input the available GPU resources $G$, the per-GPU memory budget $B$, and the SLO bound $S$. The output is an execution configuration $x$ for serving $r$ on $G$. This configuration specifies parallelism, concurrency, and memory control, including where applicable mechanisms are applied and at what intensity.

The planner optimizes goodput under memory and latency constraints:
\[
\begin{gathered}
x^* =
\arg\max_{x\in\mathcal{X}(r,G)}
\mathrm{Goodput}(r,x) \\
\text{s.t.}\quad
\mathrm{PeakMem}(r,x) \le B,\quad
\mathrm{Latency}(r,x) \le S.
\end{gathered}
\]
Here, $\mathcal{X}(r,G)$ is the legal execution-configuration space for request template $r$ on GPU resources $G$. Because this space includes the three configuration dimensions as well as fine-grained placement and intensity choices, direct enumeration is expensive.

\subsection{Memory Analysis and Latency Profiling}

\noindent\textbf{Memory analyzer.}
Given a request template and an execution configuration, the Memory Analyzer calculates the GPU memory requirement offline. It first performs a shape-only run of the diffusion pipeline to collect tensor lifetimes and derive the active-memory trace. It then passes this trace to the Static Memory Allocator to obtain the amount of GPU memory that must be reserved for this configuration.

\noindent\textbf{Latency profiler.}
Given a request template and an execution configuration, the Latency Profiler measures the execution time of the request. Profiling a complete request can take minutes or even hours, so \system profiles the latency of the operator combination executed at each point in the trace and reuses a measurement whenever the same combination appears again. For example, many DiT blocks share the same structure; under the same execution configuration, the latency measured at one point inside one DiT block can be reused for the corresponding points in other DiT blocks. Across different configurations, regions that do not encounter OOM also execute the same operator combinations at the same points, so their profiling results can be reused as well.

\subsection{Planning Workflow}

\noindent\textbf{Stage 1: Parallelism and Batch Search.}
The planner first enumerates the parallelism and batch-size candidates (Figure~\ref{fig:planner-workflow}(a)). Given $N_{\text{GPU}}$ GPUs, \system enumerates each factor of $N_{\text{GPU}}$ as the SP degree and sets $DP=N_{\text{GPU}}/SP$. For each $(SP,DP)$ layout, it increases the batch size $BS$ from one and applies two pruning rules.

First, the memory rule removes candidates that still exceed the memory budget even after all applicable memory mitigation mechanisms are enabled. Second, the SLO rule removes candidates whose latency without additional memory mitigation already exceeds the SLO, since mitigation can only add overhead. Once a batch size is pruned under a fixed $(SP,DP)$ layout, \system stops considering larger batch sizes for that layout because larger batches usually increase both memory pressure and latency. The remaining $(SP,DP,BS)$ candidates are kept for later planning stages.

\noindent\textbf{Stage 2: Locating Where Memory Exceeds the Budget.}
For each remaining $(SP,DP,BS)$ candidate, the planner sends the corresponding request template and execution configuration to the Memory Analyzer, which returns the reserved-memory trace for that configuration. The planner then scans this trace and marks each contiguous range whose reserved memory exceeds the GPU budget (Figure~\ref{fig:planner-workflow}(b)). Each marked range records the corresponding portion of the execution trace and the operators executed there, indicating where memory mitigation may be needed.

\noindent\textbf{Stage 3: Memory Mitigation Search.}
For each remaining $(SP,DP,BS)$ candidate, the planner processes all ranges whose reserved memory exceeds the GPU budget. For each range, it examines the operator combinations executed there and identifies the memory mitigation mechanisms applicable to those operators. The planner first tries mitigation choices that do not increase measured latency, such as offloading transfers that can be fully hidden by computation or chunking that keeps the operator throughput unchanged. If these choices are not sufficient, it enumerates the remaining combinations of applicable mechanisms and intensities, determines which combinations can reduce reserved memory below the budget, and uses the Latency Profiler to choose the lowest-latency feasible combination for that range (Figure~\ref{fig:planner-workflow}(c)). After all such ranges are processed, the selected mitigation choices are combined with the candidate's parallelism and batch size, producing a complete execution configuration for that candidate.

\noindent\textbf{Stage 4: Selecting the Best Plan.}
After Stage 3, each remaining $(SP,DP,BS)$ candidate has a complete execution configuration. The planner uses the profiling results to compute the end-to-end latency and goodput of each configuration, filters out configurations whose latency exceeds the SLO, and selects the remaining configuration with the highest goodput as the deployment plan (Figure~\ref{fig:planner-workflow}(d)).

\section{Implementation}
\label{sec:impl}

\textbf{Framework integration.}
\system is built on Diffusers and uses xDiT as the distributed execution backend for SP, DP, and HSDP. The offline planner emits a plan table indexed by request template and hardware configuration, which the runtime uses to launch the planned execution.

\textbf{Memory mitigation primitives.}
\system implements a unified planner interface for memory mitigation primitives, each parameterized by target location and intensity. For activation memory, \system supports activation offloading for long-lived VAE feature caches and cross-step DiT-cache activations, using asynchronous CPU-GPU transfer hooks. For weight memory, \system extends layerwise offloading and xDiT HSDP with per-layer or per-block control, so the planner can keep only selected weights on CPU or shard only selected blocks. For concurrency and activation peaks, \system supports temporary serialization, which splits only planned logical steps into sub-batches and rejoins them afterward. \system also supports chunking and fusion for selected operators: for example, it chunks DiT attention over planned dimensions and fuses selected operators around VAE upsampling, with chunk sizes and fusion locations controlled by the plan.

\textbf{Static memory runtime.}
 \system implements the static allocator with \texttt{CUDAPluggableAllocator}. Offline analysis produces a mapping from tensor identifiers to planned offsets; at runtime, allocation becomes a table lookup. A \texttt{torch.cuda.}\allowbreak\texttt{MemPool} routes planned tensor allocations to this allocator, while memory usage that cannot be derived offline like CUDA contexts and NCCL buffers remain on PyTorch's default allocator.
\section{Evaluation}
\label{sec:eval}

\subsection{Experiment Setup}
\label{sec:eval:method}

\noindent\textbf{Hardware and models.}
We run experiments on a single server with 8 NVIDIA RTX A6000 GPUs, each with 48GB memory. The machine has two NUMA domains; GPUs are connected through PCIe 4.0, with four NVLink-connected GPU pairs. We evaluate three production diffusion models: Flux.2, a 32B-parameter image generation model; CogVideoX-5B, a 5B-parameter video generation model; and LTX-2, a 13B-parameter video generation model.

\noindent\textbf{Baselines.}
We evaluate \system against three baselines:
\begin{itemize}
\item \textbf{vLLM-Omni.} Uses vLLM-Omni, a state-of-the-art diffusion serving system, and selects its best configuration by grid search.
\item \textbf{vLLM-Omni + ExtendedLib.} Adds the extended memory-control primitives introduced by \system to vLLM-Omni and selects the best configuration by grid search. However, each primitive is applied globally rather than at fine granularity.
\item \textbf{Heuristic.} Uses the extended memory-control primitives and fine-grained execution support, but greedily searches parallelism, memory control, and concurrency.
\end{itemize}
\system uses the extended memory-control primitives and jointly searches parallelism, concurrency, and memory control.

\noindent\textbf{Workloads.}
We generate request streams using a Poisson arrival process~\cite{mixfusion}. The request mix covers 21 request templates: Flux.2 uses six image resolutions, $256\times 256$, $512\times 512$, $768\times 768$, $1024\times 1024$, $1280\times 1280$, and $2048\times 2048$; CogVideoX-5B uses 720p, 900p, and 1080p, each with 49, 65, or 81 frames; and LTX-2 uses 480p, 720p, and 1080p, each with 49, 65, or 81 frames. The default arrival rates are 5.0 reqs/min for Flux.2, 7.0 reqs/min for LTX-2, and 0.2 reqs/min for CogVideoX-5B. Following prior work~\cite{tridentserve}, we set the base SLO of each request template to $2.5\times$ the latency of its best feasible strategy in the shared candidate space; the default SLO scale is 1.0 for all models.

\subsection{End-to-End Performance}
\label{sec:eval:e2e}

\begin{figure}[t]
\centering
\includegraphics[width=\linewidth]{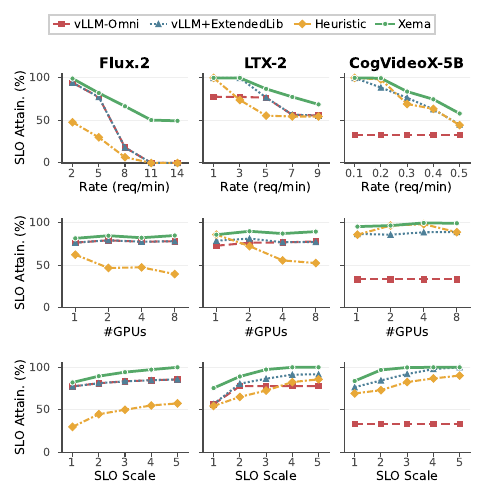}
\caption{End-to-end SLO attainment under different arrival rates, GPU counts, and SLO scales.}
\label{fig:exp1-main}
\end{figure}

\noindent\textbf{SLO attainment vs. arrival rate.}
The first row of Figure~\ref{fig:exp1-main} varies the request arrival rate on four GPUs. \system achieves the largest improvement on Flux.2 under high load: at 8 req/min, \system reaches 66.6\% SLO attainment, compared with 18.2\% for vLLM-Omni and vLLM-Omni + ExtendedLib, giving the 3.7$\times$ improvement reported in the introduction. At 11 and 14 req/min, all baselines drop to 0\% attainment, while \system still reaches 50.4\% and 49.5\%, respectively. This gap comes from the combination of batching and selective memory mitigation: Flux.2 benefits from concurrency, but coarse or poorly planned mitigation quickly turns the extra latency into SLO violations.

For video workloads, \system also maintains higher attainment as requests become more memory intensive. On LTX-2, \system reaches 68.8\% attainment at 9 req/min, while the best baseline reaches 55.8\%. On CogVideoX-5B, vLLM-Omni stays at 33.3\% across all arrival rates because several request templates remain infeasible under its native memory controls. Adding more controls improves feasibility, but global mitigation and greedy planning still lose attainment at high load: at 0.5 req/min, \system reaches 58.1\%, compared with 45.1\% for vLLM-Omni + ExtendedLib and 44.4\% for Heuristic.

\noindent\textbf{SLO attainment vs. GPU count.}
The second row of Figure~\ref{fig:exp1-main} varies the number of GPUs while scaling the arrival rate proportionally. \system remains stable as the deployment scales because the planner can make each request template feasible and then exploit additional GPUs mainly through DP. On CogVideoX-5B, \system maintains 95.3\%--99.6\% attainment from 1 to 8 GPUs, while vLLM-Omni remains fixed at 33.3\% because adding GPUs does not resolve the per-template memory infeasibility. On LTX-2, \system improves from 85.8\% on one GPU to 89.4\% on eight GPUs, whereas Heuristic drops from 85.9\% to 52.4\%. This drop indicates that greedy choices can select unnecessary SP or HSDP, whose communication overhead becomes more costly as the system scales.

\noindent\textbf{SLO attainment vs. SLO scale.}
The third row of Figure~\ref{fig:exp1-main} varies the SLO scale on four GPUs. \system's advantage is most visible under tight SLOs, where extra communication, serialization, or global mitigation directly causes deadline misses. At SLO scale 1.0, \system reaches 82.2\%, 75.6\%, and 83.9\% attainment on Flux.2, LTX-2, and CogVideoX-5B, respectively. In contrast, Heuristic reaches only 30.0\%, 54.6\%, and 69.3\% on the same workloads. As the SLO scale increases, all feasible systems improve, but \system reaches near-saturated attainment earlier: it reaches 97.3\% on Flux.2 at scale 4.0 and 100\% on LTX-2 and CogVideoX-5B at scale 4.0, while vLLM-Omni remains limited on CogVideoX-5B at 33.3\% due to OOM rather than latency.

\begin{figure}[t]
\centering
\includegraphics[width=\linewidth]{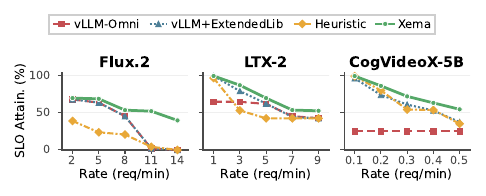}
\caption{SLO attainment under a skewed request mix.}
\label{fig:exp1-skewed}
\end{figure}

\noindent\textbf{Skewed request mix.}
Figure~\ref{fig:exp1-skewed} evaluates a skewed request mix on four GPUs. Following TetriServe~\cite{tetriserve}, each template $i$ is sampled with probability $p_i \propto \exp(\alpha L_i/L_{\max})$, where $L_i$ is its latent length, $L_{\max}$ is the maximum latent length, and larger $\alpha$ biases the mix toward larger requests; we set $\alpha=1.0$. This setting increases the fraction of memory-intensive requests and therefore stresses both feasibility and mitigation overhead. Under this skew, \system still maintains 39.9\% attainment for Flux.2 at 14 req/min, while all baselines fall to 0\%. On CogVideoX-5B, \system reaches 54.9\% at 0.5 req/min, compared with 25.0\% for vLLM-Omni, 37.7\% for vLLM-Omni + ExtendedLib, and 35.4\% for Heuristic. On LTX-2, \system reaches 52.9\% at 9 req/min, about 10 percentage points higher than all baselines. These results show that \system's gains persist when the workload shifts toward larger templates, where both fine-grained mitigation and joint configuration planning become more important.

\subsection{Extreme Workloads}
\label{sec:eval:extreme}

\begin{figure}[t]
\centering
\includegraphics[width=\linewidth]{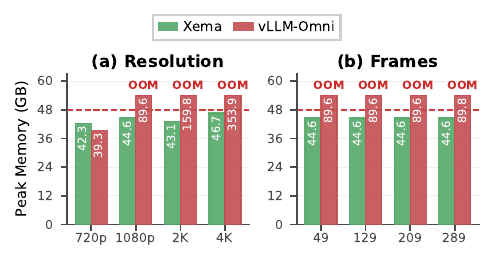}
\caption{Peak memory when scaling CogVideoX-5B request templates on one A6000 GPU.}
\label{fig:exp2}
\end{figure}

This section evaluates extreme single-request feasibility on one 48GB A6000 GPU using CogVideoX-5B, whose memory demand grows most rapidly with resolution and frame count among our evaluated models. Figure~\ref{fig:exp2} shows two sweeps: increasing resolution at 49 frames and increasing frame count at 1080p. vLLM-Omni can only complete 720p$\times$49f and OOMs for all larger resolution or frame-count settings. In contrast, \system combines activation offloading, fusion, and chunking to reduce intermediate peaks, allowing all tested requests up to 4K$\times$49f and 1080p$\times$289f to fit within the single-GPU budget. The near-flat peak memory in the frame-count sweep comes from the VAE-dominated peak, whose size is determined mainly by spatial resolution rather than frame count.

\subsection{Memory Layout Comparison}
\label{sec:eval:layout}

\begin{figure}[t]
\centering
\includegraphics[width=\linewidth]{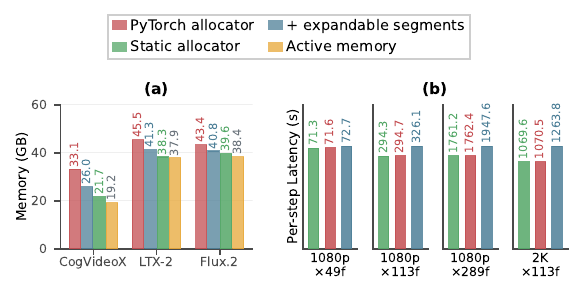}
\caption{Memory usage and per-step latency of different allocators.}
\label{fig:eval-layout}
\end{figure}

Figure~\ref{fig:eval-layout} compares \system's static allocator with PyTorch's caching allocator and PyTorch with \texttt{expandable\_segments}. Figure~\ref{fig:eval-layout}(a) shows that PyTorch reserves 33.1GB, 45.5GB, and 43.4GB on CogVideoX-5B, LTX-2, and Flux.2, while the corresponding active peaks are only 19.2GB, 37.9GB, and 38.4GB. \texttt{expandable\_segments} reduces this allocator overhead, but still leaves 26.0GB, 41.3GB, and 40.8GB reserved. In contrast, \system keeps reserved memory close to the active peak, reducing the reserved peak to 21.7GB, 38.3GB, and 39.6GB. This improves effective memory utilization from 58\% to 88\% on CogVideoX-5B and from 83\%--89\% to 97\%--99\% on LTX-2 and Flux.2. Figure~\ref{fig:eval-layout}(b) shows that this reduction does not add runtime overhead. The static allocator matches PyTorch's latency across the tested CogVideoX-5B requests because runtime allocation becomes a table lookup from tensor identifiers to preplanned addresses. By contrast, \texttt{expandable\_segments} introduces page-level mapping overhead on larger requests: for 2K$\times$113f, latency increases from 1069.6s with \system to 1263.8s with \texttt{expandable\_segments}.

\begin{figure}[t]
\centering
\includegraphics[width=\linewidth]{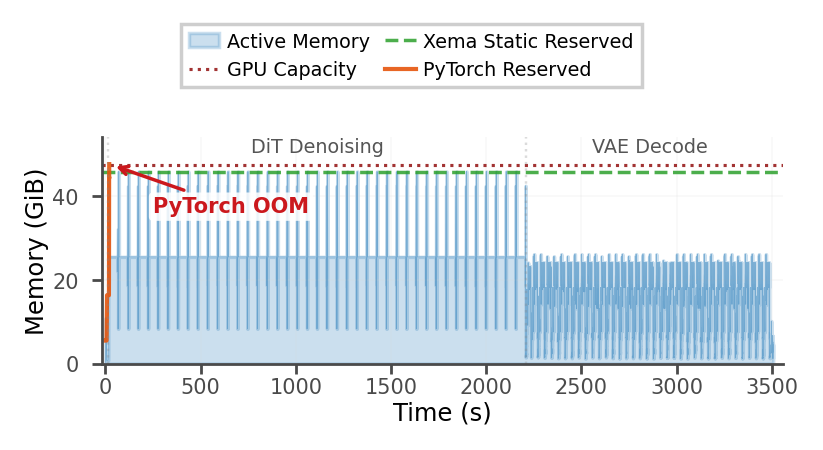}
\caption{Memory trace of CogVideoX-5B 1080p$\times$289f on one A6000 GPU.}
\label{fig:exp3-oom}
\end{figure}

\noindent\textbf{Memory trace.}
Figure~\ref{fig:exp3-oom} shows why controlling fragmentation matters near the GPU memory limit. For CogVideoX-5B 1080p$\times$289f on one A6000, PyTorch's reserved memory exceeds the 48GB budget before DiT denoising begins, even though active memory still fits. \system's static layout keeps reserved memory close to active memory, allowing the same request to complete both DiT denoising and VAE decoding.

\subsection{Planning Cost}
\label{sec:eval:planner}

\begin{table}[t]
\centering
\caption{Planning time breakdown.}
\label{tab:exp4}
\begin{tabular}{lrr}
\toprule
Step & \system & Grid Search \\
\midrule
Lifecycle analysis & 12.6 s & N/A \\
Pruning & 1.1 ms & N/A \\
Profiling & 184.4 s & 6.3 hours \\
\midrule
\textbf{Total} & \textbf{197.0 s} & \textbf{6.3 hours} \\
\bottomrule
\end{tabular}
\end{table}

Table~\ref{tab:exp4} compares the planning cost of \system and exhaustive grid search for CogVideoX-5B 1080p$\times$49f. Grid search must enumerate the full combination space of parallelism, concurrency, and memory control, and therefore profiles many infeasible or unnecessary configurations, taking 6.3 hours in total. \system first performs lifecycle analysis to obtain the memory trace, then prunes candidates that are clearly infeasible or already violate the SLO. It further restricts memory-mitigation search to the regions where memory may exceed the budget. As a result, \system reduces profiling time to 184.4 seconds and finishes planning in 197.0 seconds, two orders of magnitude faster than grid search.

\subsection{Configuration Ablation}
\label{sec:eval:ablation}

\begin{figure}[t]
\centering
\includegraphics[width=\linewidth]{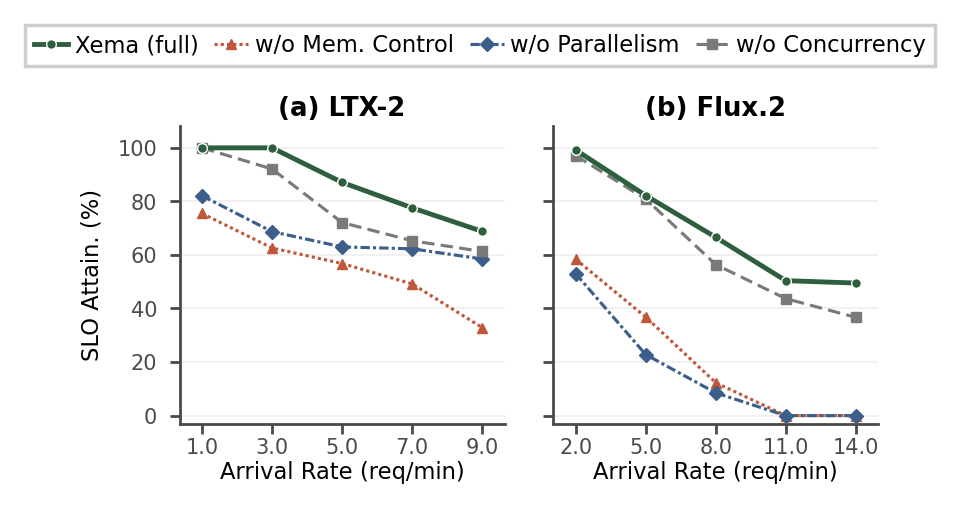}
\caption{Configuration ablation on LTX-2 and Flux.2.}
\label{fig:exp5}
\end{figure}

Figure~\ref{fig:exp5} ablates the three planning dimensions on LTX-2 and Flux.2 using restricted variants. \textit{w/o Mem. Control} removes memory-control planning by enabling all memory-control mechanisms globally, \textit{w/o Parallelism} fixes SP and HSDP to their maximum settings, and \textit{w/o Concurrency} fixes batch size to one. Each variant therefore keeps the system feasible, but forces one dimension to a conservative choice instead of letting the planner select it per request template.

The results show that all three dimensions affect SLO attainment, but their impact depends on the workload. On LTX-2 at 9 req/min, \system reaches 68.8\% attainment, compared with 32.9\% without memory-control planning, 58.5\% without parallelism planning, and 61.3\% without concurrency planning. The large drop without memory-control planning shows that globally enabling mitigation pays unnecessary latency overhead even when only localized intervals need memory reduction. On Flux.2, the interaction is more pronounced: at 8 req/min, \system reaches 66.6\%, while \textit{w/o Mem. Control} and \textit{w/o Parallelism} drop to 12.3\% and 8.5\%, respectively; at 11 req/min, both variants fall to 0\% while \system still reaches 50.4\%. This is because Flux.2 needs batching for GPU utilization, but excessive HSDP/SP or global memory controls add enough latency to turn feasible requests into SLO violations. Fixing batch size to one is less catastrophic, but still reduces attainment from 49.5\% to 36.7\% at 14 req/min. These results confirm that memory control, parallelism, and concurrency must be planned jointly rather than independently fixed to conservative settings.

\section{Related Work}
\label{sec:related}

\noindent\textbf{Diffusion model serving.}
Recent systems optimize diffusion serving through batching, model selection, adapter and image-editing workflows, patch-level or stage-level scheduling, adaptive sequence parallelism, and multimodal serving runtimes~\cite{diffserve,swiftdiffusion,katz,flashps,mixfusion,tridentserve,tetriserve,vllmomni,sglangdiffusion}. They do not jointly plan parallelism, concurrency, fine-grained memory mitigation, and static memory layout for each request template under memory and SLO constraints.

\noindent\textbf{Diffusion inference acceleration.}
Other work reduces single-request generation cost using faster samplers, latent or transformer backbones, multi-GPU execution, patch-level pipelining, feature caching, or denoising-step approximation~\cite{ddpm,ddim,paradigms,ldm,dit,cogvideox,ltxvideo,distrifusion,pipefusion,deepcache,adacache,teacache,fastcache,cachequant}. These techniques are orthogonal to \system, which chooses feasible, high-goodput configurations for heterogeneous request streams.

\noindent\textbf{GPU memory management.}
Prior memory systems use recomputation, swapping, offloading, partitioning, allocator redesign, memory virtualization, or compiler/runtime buffer reuse~\cite{checkmate,dtr,delta,swapadvisor,zero,gmlake,vattention,pytorchallocator,memomalloc,safestaticmemory,stalloc,xla,tvm-usmp}. Training-oriented systems trade activation memory for recomputation or host-device movement, while KV-cache systems optimize long-lived token caches. Compiler and allocator approaches can reuse buffers or reduce fragmentation when graph or lifetime information is visible. \system instead uses predictable diffusion lifetimes to build a static runtime allocator and integrate memory layout with serving-time goodput optimization.
\section{Conclusion}
\label{sec:conclusion}

This paper presents \system, a diffusion serving system that exploits template-static tensor lifetimes to reduce the cost of memory mitigation and automatically plan execution configurations under GPU memory and SLO constraints. For each request template, \system derives memory traces offline, applies mitigation only where and by the amount needed, and uses a static memory allocator to reduce fragmentation and keep runtime allocation consistent with offline analysis. Built on these mechanisms, \system selects high-goodput configurations across memory control, parallelism, and concurrency. Compared with existing serving systems, \system improves SLO attainment on production image and video diffusion models, expands the range of feasible request templates, and reduces configuration search cost.

\bibliographystyle{ACM-Reference-Format}
\bibliography{refs}

\end{document}